\documentclass{article}
\usepackage{graphicx}
\begin{document}

\centerline{GAMOW-TELLER (GT$\pm$) STRENGTH DISTRIBUTIONS}\vskip
0.03in

\centerline{OF $^{56}Ni$ FOR GROUND AND EXCITED STATES}\vskip 0.2in

\textbf{Jameel-Un Nabi\thanks{ Corresponding author\\
e-mail: jnabi00@gmail.com}, Muneeb-Ur Rahman, Muhammad Sajjad}\vskip
0.2in \centerline{Faculty of Engineering Sciences, Ghulam Ishaq Khan
Institute of Engineering} \centerline{Sciences and Technology, Topi
23640, Swabi, NWFP, Pakistan}\vskip0.1in Gamow-Teller (GT)
transitions play an important and consequential role in many
astrophysical phenomena. These include, but are not limited to,
electron and positron capture rates which determine the fate of
massive stars and play an intricate role in the dynamics of core
collapse. These $GT_{\pm}$ transitions rates are the significant
inputs in the description of supernova explosions. $GT_{\pm}$
strength function values are sensitive to the $^{56}Ni$ core
excitation in the middle \textit{pf}-shell region and to the size of
the model space as well. We used the pn-QRPA theory for extracting
the GT strength for ground and excited states of $^{56}Ni$. We then
used these GT strength distributions to calculate the electron
\textit{and} positron capture rates which show differences with the
earlier calculations. One curious finding of this paper is our
enhanced electron capture rates on $^{56}Ni$ at presupernova
temperatures. These differences need to be taken into account for
the modeling of the early stages of Type II supernova evolution.
\vskip 0.2in

\begin{center}{\Large\textbf{1. Introduction}}\end{center}

Weak interactions play a conclusive role in the evolution of massive
stars at the presupernova stage and supernova explosions. These
explosions mark the end of the life of massive stars. The massive
stars consist of concentric shells which are the relics of their
previous burning phases. The helium burning shell continues to add
ashes to the carbon-oxygen core. This results in the contraction of
the core and eventually initiates the carbon burning which then
leads to a variety of by-products, such as ${}_{8}^{16} {\it O}$,
${}_{10}^{20} Ne$, $_{11}^{23} Na$, $_{12}^{23} Mg$, and $_{12}^{24}
Mg$ [1]. What follows is a succession of nuclear reaction sequences
which depend sensitively on the mass of the star. When each reaction
sequence reaches equilibrium, an ``onion-like'' shell structure
develops in the interior of the star.

Stars with initial mass about 10$M_{\odot } $ or more ignite carbon
in the core non-degenerately [2]. Owing to neutrino (and
antineutrino) emission at the high temperatures involved, due to
$e^{\pm } $ annihilation and other processes, subsequent evolution
is greatly accelerated. The nuclear time-scale becomes shorter than
the thermal one because carbon, oxygen, and silicon burning produce
nuclei with masses progressively nearer the iron peak of the binding
energy curve, and consequently less and less energy is generated per
gram of fuel.

When the core attains high density and temperature, the photons
having enough energy destroy heavy nuclei; a process known as
photodisintegration. In a very short span, this photodisintegration
reverses what the star has been trying to do its entire life, i.e.
to produce more massive elements than hydrogen and helium. This
stripping down of iron to individual protons and neutrons is highly
endothermic. This saps the thermal energy from the gas that would
otherwise have resulted in the pressure necessary to support the
core of the star.

At high temperature and density, the electrons supporting the star
through degeneracy pressure are eaten up by heavy nuclei and protons
that were produced during photodisintegration process, and thus lead
to the neutronization of star. Electron capture and
photodisintegration cost the core energy, reduce its electron
density and this results in an accelerated core collapse. The
collapse is very sensitive to entropy and to the number of lepton to
baryon ratio, \textit{$Y_{e}$}. These two quantities are mainly
determined by weak interaction processes. In the inner region of the
core, this collapse is homologous and subsonic having velocity of
the collapse proportional to the distance away from the center of
the star, while the outer regions collapse supersonically [3].

The structure of the progenitor star, including that of its core,
plays a pivotal role in the development of the explosion process.
Electron capture reduces the number of electrons available for
pressure support. At higher densities, $\rho  \approx
10^{11}g/cm^{3}$, electron capture produces neutrinos which escape
the star carrying away energy and entropy from the core. Electron
capture during the final evolution of a massive star is dominated
by Fermi and Gamow-Teller (GT) transitions. The energies of the
electrons are high enough to induce transitions to the GT
resonance.  The electron capture rates are very sensitive to the
distribution of the $GT_{+}$ strength (in this direction a proton
is changed into a neutron).

Bethe et al. [4] showed that, as a result of electron capture, the
average number of nucleons per nucleus ($\overline{A}$) moves
upward. Nevertheless, we can say that there is tendency for
$\overline{A}$  to increase with decreasing \textit{$Y_{e}$}. During
collapse, the entropy of the core decides whether electron capture
occur on heavy nuclei or on free proton (produced during
photodisintegration). The total entropy of the stellar core is the
sum of the entropies due to nuclear excitation and that of the free
nucleons. At low entropies $(S/k_{B}\approx 1)$ captures on heavy
nuclei dominate the total rate. These entropies of the stellar core
do occur for the star of main sequence mass between 10 and 25
$M_{\odot }$ and density range $10^{9}$- $10^{12} g/cm^{3}$ [5].

Electron captures on proton and positron captures on neutron play a
very crucial role in the supernovae dynamics. During the collapse
and accretion phases, these processes exhaust electrons, thus
decreasing the degenerate pressure of electrons in the stellar core.
Meanwhile, they produce neutrinos which carry the binding energy
away. Therefore, electron and positron captures play key role in the
dynamics of the formation of bounce shock of supernova. The Type II
supernovae take place in heavy stars. The positron captures are of
great importance in high temperature and low density locations. In
such conditions, a rather high concentration of positron can be
reached from $e^{-} +e^{+} \leftrightarrow \gamma +\gamma $
equilibrium favoring the $e^{-} e^{+} $ pairs. The electron captures
on proton and positron captures on neutron are considered important
ingredients in the modeling of Type II supernovae [6].

Proton-neutron quasi particle random phase approximation (pn-QRPA)
theory and shell model are extensively used for the calculations of
capture rates in the stellar environment. Each model has its own
associated pros and cons. Shell model lays more emphasis on
interaction of nucleons as compared to correlations whereas pn-QRPA
puts more weight on correlations. One big advantage of using pn-QRPA
theory is that it gives us the liberty of performing calculations in
a luxurious model space (up to $7\hbar\omega$). The pn-QRPA method
considers the residual correlations among the nucleons via one
particle one hole (1p-1h) excitations in a large model spaces. The
authors in [7] extended the QRPA model to configurations more
complex than (1p-1h). The pn-QRPA formalism was successfully
employed to calculate weak interaction rates for 178 sd-shell [7]
and 650 fp/fpg-shell [8] nuclide in stellar matter. Later the decay
and capture rates of nuclei of astrophysical importance were studied
separately in detail and were compared with earlier calculations
wherever possible both in sd-shell [9] and fp-shell (e.g. [10, 11])
regions.

Knowing the importance of the electron and positron capture
processes in the evolution of stars many authors estimated these
rates independently employing different models. Fuller et al.
(referred as FFN) [12] estimated these rates for the nuclei in the
mass range A = 45 - 60. They related these capture processes to the
GT resonance. Aufderheide et al. [13, 14] then updated the rates of
FFN and compiled a list of important nuclide and showed that these
nuclide strongly affect \textit{$Y_{e}$} via the electron capture
processes.  They ranked $^{56}Ni$ amongst the top ten nuclei which
play a vital role in the deleptonization of the core. This isotope
of nickel is abundant in the presupernova environment, and is
considered to be a dominant role player among other iron-regime
nuclei in the evolution of stellar core. The GT response is
astrophysically important for a number of nuclide, particularly
$^{56}Ni$.

Recently the calculations of electron capture rates on $^{55}Co$ and
$^{56}Ni$ using the pn-QRPA theory were presented and compared with
earlier calculations [15]. There the authors also discussed the
possible applications of these calculated rates in astrophysical
environments. In this paper we present for the first time the GT
strength distributions (both plus and minus) from the parent
\textit{and} excited states of $^{56}Ni$. We also present the
associated electron and positron capture rates for this important
isotope of nickel. Comparison with earlier calculations wherever
possible is also being presented. We used the pn-QRPA model to
generate GT strength distributions and performed state by state
calculations of the associated electron and positron capture rates.
These calculated rates were summed over all parent and daughter
states until satisfactory convergence was achieved.

We made the following assumptions to calculate electron and
positron capture rates on $^{56}Ni$.

\begin{enumerate}
\item Forbidden transitions were not taken into account. Only the
allowed Gamow-Teller and superallowed Fermi transitions were
calculated.\item Electrons and positrons, in stellar matter, were
assumed to follow the energy distribution of a Fermi gas.\item Fermi
functions were used in the phase space integrals to represent the
distortion of electron (positron) wavefunctions (due to coulombic
interactions of these with the nucleus).

\item  Neutrinos and antineutrinos which are produced were assumed
to escape freely from the core without interacting with any
particle. We neglected the capture of (anti) neutrinos in our
calculations.
\end{enumerate}

\begin{center}{\Large\textbf{2. General Formalism}}\end{center}

In this paper we present the calculated capture rates on $^{56}Ni$
for the following two processes mediated by charge weak
interaction:

\begin{enumerate}
\item  Electron capture

\[{}_{Z}^{A} X\, +\, e^{-} \to \, {}_{Z-1}^{A} X\, +\, \nu .\]
\item  Positron capture

\[{}_{Z}^{A} X\, +\, e^{+} \, \to \, {}_{Z+1}^{A} X\, +\, \bar{\nu }.\]
\end{enumerate}

These processes play an important role in the evolution of
presupernova core. To calculate these electron capture and positron
capture rates in the stellar environment, we used the following
formalism.

The Hamiltonian of our model was chosen as

\begin{equation} \label{GrindEQ__1_} {\rm H}^{{\rm QRPA}} {\rm \; =\; H}^{{\rm sp}} {\rm \; +\; V}^{{\rm pair}} {\rm \; +\; V}_{{\rm GT}}^{{\rm ph}} {\rm \; +\; V}_{{\rm GT}}^{{\rm pp}} . \end{equation}

Here $H^{sp} $ is the single-particle Hamiltonian, $V^{pair} $is the
pairing force, $V_{GT}^{ph} $ is the particle-hole (ph) Gamow-Teller
force, and $V_{GT}^{pp} $ is the particle-particle (pp) Gamow-Teller
force. Wave functions and single particle energies were calculated
in the Nilsson model [16], which takes into account the nuclear
deformations.  Pairing was treated in the BCS approximation. The
proton-neutron residual interactions occur in two different forms,
namely as particle-hole and particle-particle interaction. The
interactions were given separable form and were characterized by two
interaction constants $\chi $ (characterizing the particle-hole
force) and $\kappa $ (characterizing the particle-particle force).
The selections of these two constants were done in an optimal
fashion. For details of the fine tuning of the Gamow-Teller strength
parameters, we refer to [17, 18]. In this work, we took the values
of $\chi $ = 0.5 MeV and $\kappa $ = 0.065 MeV for $^{56}Ni$.

Other parameters required for the calculation of capture rates are
the Nilsson potential parameters, the deformation, the pairing gaps,
and the Q-value of the reaction. Nilsson-potential parameters were
taken from [19] and the Nilsson oscillator constant was chosen as
$\hbar \omega =41A^{-1/3} (MeV)$, the same for protons and neutrons.
The calculated half-lives depend only weakly on the values of the
pairing gaps [20]. Thus, the traditional choice of $\Delta _{p}
=\Delta _{n} =12/\sqrt{A} (MeV)$ was applied in the present work.
For details regarding the QRPA wave functions and calculation of
weak rates we refer to [11]. Q-values were taken from the recent
mass compilation of Audi et al. [21].

The Fermi operator is independent of space and spin, and as a
result the Fermi strength is concentrated in a very narrow
resonance centered around the isobaric analogue state (IAS) for
the ground and excited states. The energy of the IAS was
calculated according to the prescription given in [22,
pp.111-112], whereas the reduced transition probability is given
by

\[B(F)=T(T+1)-T_{zi} T_{zf} ,\]
where $T$ is the nuclear isospin, and $T_{zi} $, $T_{zf} $ are the
third components of the isospin of initial and final analogue
states, respectively.

The parent excited states can be constructed as phonon-correlated
multi-quasiparticles states. The transition amplitudes between the
multi-quasiparticle states can be reduced to those of
single-particle states. Excited states of an even-even nucleus are
two-proton quasiparticle states and two-neutron quasiparticle
states. Transitions from these initial states are possible to final
proton-neutron quasiparticles pair states in the odd-odd daughter
nucleus. The transition amplitudes and their reduction to correlated
(c) one-quasiparticle states are given by

\[\left\langle p^{f} n_{c}^{f} \left|t_{\pm } \sigma _{-\mu } \right|\right. \left. p_{1}^{i} p_{2c}^{i} \right\rangle \, =\, -\delta (p^{f,} p_{2}^{i} )\left\langle n_{c}^{f} \left|t_{\pm } \sigma _{-\mu } \right|\right. \left. p_{1c}^{i} \right\rangle \, +\, \delta (p^{f,} p_{1}^{i} )\left\langle n_{c}^{f} \left|t_{\pm } \sigma _{-\mu } \right|\right. \left. p_{2c}^{i} \right\rangle .\, \, \, \, \, \, \, (2)\]

\[\left\langle p^{f} n_{c}^{f} \left|t_{\pm } \sigma _{\mu } \right|\right. \left. n_{1}^{i} n_{2c}^{i} \right\rangle \, =\, +\, \delta (n^{f,} n_{2}^{i} )\left\langle p_{c}^{f} \left|t_{\pm } \sigma _{\mu } \right|\right. \left. n_{1c}^{i} \right\rangle \, -\, \delta (n^{f,} n_{1}^{i} )\left\langle p_{c}^{f} \left|t_{\pm } \sigma _{\mu } \right|\right. \left. n_{2c}^{i} \right\rangle .\, \, \, \, \, \, \, \, (3)\]
Here $\mu $= -1, 0, 1, are the spherical components of the spin
operator.

States in an odd-odd nucleus are expressed in quasiparticle
transformation by two-quasiparticle states (proton-neutron pair
states) or by four-quasiparticle states (three-proton, one-neutron
or one-proton three-neutron quasiparticle states). The reduction of
two-quasiparticle states to correlated (c) one-quasiparticle states
is given by

\[\left\langle p_{1}^{f} p_{2c}^{f} \left|t_{\pm } \sigma _{\mu } \right|\right. \left. p^{i} n_{c}^{i} \right\rangle =\delta (p_{1}^{f} ,p^{i} )\left\langle p_{2c}^{f} \left|t_{\pm } \sigma _{\mu } \right|\right. \left. n_{c}^{i} \right\rangle \, -\, \delta (p_{2}^{f} ,p^{i} )\left\langle p_{1c}^{f} \left|t_{\pm } \sigma _{\mu } \right|\right. \left. n_{c}^{i} \right\rangle .\, \, \, \, \, \, \, \, (4)\]

\[\left\langle n_{1}^{f} n_{2c}^{f} \left|t_{\pm } \sigma _{-\mu } \right|\right. \left. p^{i} n_{c}^{i} \right\rangle \, =\, \delta (n_{2}^{f} ,n^{i} )\left\langle n_{1c}^{f} \left|t_{\pm } \sigma _{-\mu } \right|\right. \left. p_{c}^{i} \right\rangle \, -\, \delta (n_{1}^{f} ,n^{i} )\left\langle n_{2c}^{f} \left|t_{\pm } \sigma _{-\mu } \right|\right. \left. p_{c}^{i} \right\rangle .\, \, \, \, \, \, \, \, \, \, (5)\]

While four-quasiparticle states are simplified as

\[\begin{array}{l} {\left\langle p_{1}^{f} p_{2}^{f} n_{1}^{f} n_{2c}^{f} \left|t_{\pm } \sigma _{-\mu } \right|\right. \left. p_{1}^{i} p_{2}^{i} p_{3}^{i} n_{1c}^{i} \right\rangle \, =\, \delta (n_{2}^{f} ,n_{1}^{i} )\left[\delta (p_{1}^{f} ,p_{2}^{i} )\delta (p_{2}^{f} ,p_{3}^{i} )\left\langle n_{1c}^{f} \left|t_{\pm } \sigma _{-\mu } \right|\right. \left. p_{1c}^{i} \right\rangle \right. } \\ {\, \, \, \, \, \, \, \, \, \, \, \, \, \, \, \, \, \, \, \, \, \, \, \, \, \, \, \, \, \, \, \, \, \, \, \, \, \, \, \, \, \, \, \, \, \, \, \, \, \, \, \, \, \, \, \, \, \, \, \, \, \, \, \, \, -\delta (p_{1}^{f} ,p_{1}^{i} )\delta (p_{2}^{f} ,p_{3}^{i} )\left\langle n_{1c}^{f} \left|t_{\pm } \sigma _{-\mu } \right|\right. \left. p_{2c}^{i} \right\rangle } \\ {\, \, \, \, \, \, \, \, \, \, \, \, \, \, \, \, \, \, \, \, \, \, \, \, \, \, \, \, \, \, \, \, \, \, \, \, \, \, \, \, \, \, \, \, \, \, \, \, \, \, \, \, \, \, \, \, \, \, \, \, \, \, \, \, \, \left. +\, \delta (p_{1}^{f} ,p_{1}^{i} )\delta (p_{2}^{f} ,p_{2}^{i} )\left\langle n_{1c}^{f} \left|t_{\pm } \sigma _{-\mu } \right|\right. \left. p_{3c}^{i} \right\rangle \right]} \\ {\, \, \, \, \, \, \, \, \, \, \, \, \, \, \, \, \, \, \, \, \, \, \, \, \, \, \, \, \, \, \, \, \, \, \, \, \, \, \, \, \, \, \, \, \, \, \, \, \, \, \, \, \, \, \, \, \, \, \, \, \, \, \, \, \, -\, \delta (n_{1}^{f} ,n_{1}^{i} )\left[\delta (p_{1}^{f} ,p_{2}^{i} )\delta (p_{2}^{f} ,p_{3}^{i} )\right. \left\langle n_{2c}^{f} \left|t_{\pm } \sigma _{-\mu } \right|\right. \left. p_{1c}^{i} \right\rangle } \\ {\, \, \, \, \, \, \, \, \, \, \, \, \, \, \, \, \, \, \, \, \, \, \, \, \, \, \, \, \, \, \, \, \, \, \, \, \, \, \, \, \, \, \, \, \, \, \, \, \, \, \, \, \, \, \, \, \, \, \, \, \, \, \, \, \, -\delta (p_{1}^{f} ,p_{1}^{i} )\delta (p_{2}^{f} ,p_{3}^{i} )\left\langle n_{2c}^{f} \left|t_{\pm } \sigma _{-\mu } \right|\right. \left. p_{2c}^{i} \right\rangle } \\ {\, \, \, \, \, \, \, \, \, \, \, \, \, \, \, \, \, \, \, \, \, \, \, \, \, \, \, \, \, \, \, \, \, \, \, \, \, \, \, \, \, \, \, \, \, \, \, \, \, \, \, \, \, \, \, \, \, \, \, \, \, \, \, \, \, +\, \delta (p_{1}^{f} ,p_{1}^{i} )\delta (p_{2}^{f} ,p_{2}^{i} )\left. \left\langle n_{2c}^{f} \left|t_{\pm } \sigma _{-\mu } \right|\right. \left. p_{3c}^{i} \right\rangle \right].\, \, \, \, \, \, \, \, \, \, \, \, \, \, \, \, \, \, \, \, \, (6)} \end{array}\]

\[\begin{array}{l} {\left\langle p_{1}^{f} p_{2}^{f} p_{3}^{f} p_{4c}^{f} \left|t_{\pm } \sigma _{\mu } \right|\right. \left. p_{1}^{i} p_{2}^{i} p_{3}^{i} n_{1c}^{i} \right\rangle =\, -\, \delta (p_{2}^{f} ,p_{1}^{i} )\delta (p_{3}^{f} ,p_{2}^{i} )\delta (p_{4}^{f} ,p_{3}^{i} )\left\langle p_{1c}^{f} \left|t_{\pm } \sigma _{\mu } \right|\right. \left. n_{1c}^{i} \right\rangle } \\ {\, \, \, \, \, \, \, \, \, \, \, \, \, \, \, \, \, \, \, \, \, \, \, \, \, \, \, \, \, \, \, \, \, \, \, \, \, \, \, \, \, \, \, \, \, \, \, \, \, \, \, \, \, \, \, \, \, \, \, \, \, \, \, \, \, \, +\, \delta (p_{1}^{f} ,p_{1}^{i} )\delta (p_{3}^{f} ,p_{2}^{i} )\delta (p_{4}^{f} ,p_{3}^{i} )\left\langle p_{2c}^{f} \left|t_{\pm } \sigma _{\mu } \right|\right. \left. n_{1c}^{i} \right\rangle } \\ {\, \, \, \, \, \, \, \, \, \, \, \, \, \, \, \, \, \, \, \, \, \, \, \, \, \, \, \, \, \, \, \, \, \, \, \, \, \, \, \, \, \, \, \, \, \, \, \, \, \, \, \, \, \, \, \, \, \, \, \, \, \, \, \, \, -\, \delta (p_{1}^{f} ,p_{1}^{i} )\delta (p_{2}^{f} ,p_{2}^{i} )\delta (p_{4}^{f} ,p_{3}^{i} )\left\langle p_{3c}^{f} \left|t_{\pm } \sigma _{\mu } \right|\right. \left. n_{1c}^{i} \right\rangle } \\ {\, \, \, \, \, \, \, \, \, \, \, \, \, \, \, \, \, \, \, \, \, \, \, \, \, \, \, \, \, \, \, \, \, \, \, \, \, \, \, \, \, \, \, \, \, \, \, \, \, \, \, \, \, \, \, \, \, \, \, \, \, \, \, \, \, +\, \delta (p_{1}^{f} ,p_{1}^{i} )\delta (p_{2}^{f} ,p_{2}^{i} )\delta (p_{3}^{f} ,p_{3}^{i} )\left\langle p_{4c}^{f} \left|t_{\pm } \sigma _{\mu } \right|\right. \left. n_{1c}^{i} \right\rangle .\, \, \, \, {\rm (7)}} \end{array}\]

\[\begin{array}{l} {\left\langle p_{1}^{f} p_{2}^{f} n_{1}^{f} n_{2c}^{f} \left|t_{\pm } \sigma _{\mu } \right|\right. \left. p_{1}^{i} n_{1}^{i} n_{2}^{i} n_{3c}^{i} \right\rangle =\delta (p_{1}^{f} ,p_{1}^{i} )\left[\delta (n_{1}^{f} ,n_{2}^{i} )\delta (n_{2}^{f} ,n_{3}^{i} )\left\langle p_{2c}^{f} \left|t_{\pm } \sigma _{\mu } \right|\right. \left. n_{1c}^{i} \right\rangle \right. } \\ {\, \, \, \, \, \, \, \, \, \, \, \, \, \, \, \, \, \, \, \, \, \, \, \, \, \, \, \, \, \, \, \, \, \, \, \, \, \, \, \, \, \, \, \, \, \, \, \, \, \, \, \, \, \, \, \, \, \, \, \, \, \, \, \, -\delta (n_{1}^{f} ,n_{1}^{i} )\delta (n_{2}^{f} ,n_{3}^{i} )\left\langle p_{2c}^{f} \left|t_{\pm } \sigma _{\mu } \right|\right. \left. n_{2c}^{i} \right\rangle } \\ {\, \, \, \, \, \, \, \, \, \, \, \, \, \, \, \, \, \, \, \, \, \, \, \, \, \, \, \, \, \, \, \, \, \, \, \, \, \, \, \, \, \, \, \, \, \, \, \, \, \, \, \, \, \, \, \, \, \, \, \, \, \, \, \, +\delta (n_{1}^{f} ,n_{1}^{i} )\delta (n_{2}^{f} ,n_{2}^{i} )\left. \left\langle p_{2c}^{f} \left|t_{\pm } \sigma _{\mu } \right|\right. \left. n_{3c}^{i} \right\rangle \right]} \\ {\, \, \, \, \, \, \, \, \, \, \, \, \, \, \, \, \, \, \, \, \, \, \, \, \, \, \, \, \, \, \, \, \, \, \, \, \, \, \, \, \, \, \, \, \, \, \, \, \, \, \, \, \, \, \, \, \, \, \, \, \, \, \, -\delta (p_{2}^{f} ,p_{1}^{i} )\left[\delta (n_{1}^{f} ,n_{2}^{i} )\right. \delta (n_{2}^{f} ,n_{3}^{i} )\left\langle p_{1c}^{f} \left|t_{\pm } \sigma _{\mu } \right|\right. \left. n_{1c}^{i} \right\rangle } \\ {\, \, \, \, \, \, \, \, \, \, \, \, \, \, \, \, \, \, \, \, \, \, \, \, \, \, \, \, \, \, \, \, \, \, \, \, \, \, \, \, \, \, \, \, \, \, \, \, \, \, \, \, \, \, \, \, \, \, \, \, \, \, \, -\delta (n_{1}^{f} ,n_{1}^{i} )\delta (n_{2}^{f} ,n_{3}^{i} )\left\langle p_{1c}^{f} \left|t_{\pm } \sigma _{\mu } \right|\right. \left. n_{2c}^{i} \right\rangle } \\ {\, \, \, \, \, \, \, \, \, \, \, \, \, \, \, \, \, \, \, \, \, \, \, \, \, \, \, \, \, \, \, \, \, \, \, \, \, \, \, \, \, \, \, \, \, \, \, \, \, \, \, \, \, \, \, \, \, \, \, \, \, \, \, +\delta (n_{1}^{f} ,n_{1}^{i} )\left. \delta (n_{2}^{f} ,n_{2}^{i} )\left\langle p_{1c}^{f} \left|t_{\pm } \sigma _{\mu } \right|\right. \left. n_{3c}^{i} \right\rangle \right].\, \, \, \, \, \, \, \, \, \, \, \, \, \, \, \, \, \, (8)} \end{array}\]

\[\begin{array}{l} {\left\langle n_{1}^{f} n_{2}^{f} n_{3}^{f} n_{4c}^{f} \left|t_{\pm } \sigma _{-\mu } \right|\right. \left. p_{1}^{i} n_{1}^{i} n_{2}^{i} n_{3c}^{i} \right\rangle =\, \delta (n_{2}^{f} ,n_{1}^{i} )\delta (n_{3}^{f} ,n_{2}^{i} )\delta (n_{4}^{f} ,n_{3}^{i} )\left\langle n_{1c}^{f} \left|t_{\pm } \sigma _{-\mu } \right|\right. \left. p_{1c}^{i} \right\rangle } \\ {\, \, \, \, \, \, \, \, \, \, \, \, \, \, \, \, \, \, \, \, \, \, \, \, \, \, \, \, \, \, \, \, \, \, \, \, \, \, \, \, \, \, \, \, \, \, \, \, \, \, \, \, \, \, \, \, \, \, \, \, \, \, \, \, -\delta (n_{1}^{f} ,n_{1}^{i} )\delta (n_{3}^{f} ,n_{2}^{i} )\delta (n_{4}^{f} ,n_{3}^{i} )\left\langle n_{2c}^{f} \left|t_{\pm } \sigma _{-\mu } \right|\right. \left. p_{1c}^{i} \right\rangle } \\ {\, \, \, \, \, \, \, \, \, \, \, \, \, \, \, \, \, \, \, \, \, \, \, \, \, \, \, \, \, \, \, \, \, \, \, \, \, \, \, \, \, \, \, \, \, \, \, \, \, \, \, \, \, \, \, \, \, \, \, \, \, \, \, \, +\delta (n_{1}^{f} ,n_{1}^{i} )\delta (n_{2}^{f} ,n_{2}^{i} )\delta (n_{4}^{f} ,n_{3}^{i} )\left\langle n_{3c}^{f} \left|t_{\pm } \sigma _{-\mu } \right|\right. \left. p_{1c}^{i} \right\rangle } \\ {\, \, \, \, \, \, \, \, \, \, \, \, \, \, \, \, \, \, \, \, \, \, \, \, \, \, \, \, \, \, \, \, \, \, \, \, \, \, \, \, \, \, \, \, \, \, \, \, \, \, \, \, \, \, \, \, \, \, \, \, \, \, \, \, \, -\delta (n_{1}^{f} ,n_{1}^{i} )\delta (n_{2}^{f} ,n_{2}^{i} )\delta (n_{3}^{f} ,n_{3}^{i} )\left\langle n_{4c}^{f} \left|t_{\pm } \sigma _{-\mu } \right|\right. \left. p_{1c}^{i} \right\rangle .\, \, \, \, (9)} \end{array}\]

For all quasiparticle transition amplitudes (Eqns. (2)-(9)), we took
into account the antisymmetrization of the single-quasiparticle
states

\[\begin{array}{l} {p_{1}^{f} <p_{2}^{f} <p_{3}^{f} <p_{4}^{f} } \\ {n_{1}^{f} <n_{2}^{f} <n_{3}^{f} <n_{4}^{f} } \\ {p_{1}^{i} <p_{2}^{i} <p_{3}^{i} <p_{4}^{i} } \\ {n_{1}^{i} <n_{2}^{i} <n_{3}^{i} <n_{4}^{i} } \end{array}\]

GT transitions of phonon excitations for every excited state were
also taken into account. We also assumed that the quasiparticles in
the parent nucleus remained in the same quasiparticle orbits.

In order to further increase the reliability of our calculations, we
did incorporate experimental data wherever applicable. The
calculated excitation energies (along with their $log ft$ values, if
available) were replaced with the measured one when they were within
0.5 MeV of each other. Missing measured states were inserted.
However, we did not replace (insert) theoretical levels with the
experimental ones beyond the level in experimental compilations
without definite spin and/or parity assignment.

\begin{center}{\Large\textbf{3. Results and Discussion}}\end{center}

$^{56}Ni$ is a doubly magic nucleus which is believed to be
copiously produced in the supernova conditions and is considered to
be a prime candidate for electron capturing. In this work we
considered 30 states (up to excitation energy of 10 MeV) in
$^{56}Ni$.  States higher in energy have a negligible probability of
occupation for the temperature and density scales chosen for this
phase of collapse. Table 1 lists the calculated parent excited
states of $^{56}Ni$ in order of increasing energy. We start by
presenting the GT strength distribution functions for the ground and
first two excited states of $^{56}Ni$. Complete set of GT strength
distribution functions for higher excited states can be requested by
email to the corresponding author. We considered around 200 states
of daughters, $^{56}Co$ and $^{56}Cu$, for electron and positron
captures, respectively, up to excitation energy around 45 MeV. GT
transitions are dominant excitation mode for the electron and
positron captures during the presupernova evolution. The energy
dependence of weak interaction matrix elements (or equivalently, the
GT strength distributions) is unknown for many nuclei of potential
importance in presupernova stars and collapsing cores. The centroid
of the GT distribution determines the effective energy of electron
capture from the ground state of the parent nucleus to the excited
state of the daughter nucleus. This along with the electron-Fermi
energy determines which nuclei are able to capture electron from, or
$\beta $-decay onto the Fermi-sea at a given temperature and density
and thus control the rate at which the abundance of a particular
nuclei would change in the presupernova core. The GT strength
distributions for the electron captures and positron captures are
shown in Figs. 1 and 2, respectively. Table 2a states the
$B(GT_{+})$ strength values for the ground state of $^{56}Ni$
whereas Table 2b gives the $B(GT_{-})$ strength values. The
strengths are given up to energy of 10 MeV in daughter nuclei.
Calculated GT strength of magnitude less than $10^{-3}$ are not
included in this table. (In the online version of this paper we
replace Table 2 by Table 4 which also contains the GT strength
distribution functions for the first and second excited states of
$^{56}Ni$ in both directions.) For the calculation of the associated
electron captures on $^{56}Ni$, the authors in [23] calculated the
$B(GT_{+})$ strength only from the ground state. Our calculations of
electron capture rates include contributions from the ground as well
as the 30 excited states given in Table 1. Our calculations show
that for the ground state of $^{56}Ni$ the centroid of the GT$_{+}$
strength resides at energy around 5.7 MeV in daughter $^{56}Co$ (see
also [15]). FFN [12] placed the GT$_{+}$ resonance in $^{56}Co$  at
energy 3.8 MeV. The GT$_{+}$ centroid of [23] is at energy around
2.5-3.0 MeV in daughter $^{56}Co$. The GT$_{+}$ centroids for the
first and second excited states of $^{56}Ni$ are around 7.9 MeV and
11.4 MeV in daughter $^{56}Co$, respectively. For the ground state
of $^{56}Ni$, we calculated total GT$_{+}$ strength of 8.9 as
compared to the values 10.1 and 9.8$\pm$4 calculated by [23] and
shell model Monte Carlo calculations (SMMC) [24], respectively.

Analyzing B($GT_{-}$) strength (Fig. 2), we note that our ground
state GT centroid resides at energy around 4.7 MeV in daughter,
$^{56}Cu$. For positron captures, we calculated the total GT$_{-}$
strength for the ground state of 7.4 for $^{56}Ni$ while authors in
[25] calculated it to be 11.4 (see their Table 3, experimental
values were not mentioned). For the first and second excited states
our GT$_{-}$ centroid resides around 7.6 MeV and 8.6 MeV in daughter
$^{56}Cu$, respectively.

Fig. 3 shows the variation with densities and temperatures of our
calculated electron capture rates for $^{56}Ni$. The temperature
scale $T_{9}$ measures the temperature in $10^{9}K$ and the density
in the inset has units of $g/cm^{3}$. It is pertinent to mention
that contributions from all excited states are included in the final
calculation of these capture rates. We calculated these weak rates
for densities in the range $(10^{0.5}-10^{11})g/cm^{3}$ and for
temperature scales $T_{9} = 0.5$ to $30$. We note that the electron
capture rates increase with increasing temperatures and densities.
It is also worth mentioning that for low and intermediate densities
in the range $(10^{0.5}-10^{8})g/cm^{3}$ the electron capture rates
converge to a value of around 500 $s^{-1}$ at $T_{9} = 30$. At
higher densities order of magnitude differences start to build in
between the corresponding rates. The gradient of the curves at low
and intermediate temperatures ($T_{9} = 0.5$ to $10$) also decreases
with increasing density. At densities in the vicinity of $10^{11}
g/cm^{3}$ the electron capture rates remain constant until the
stellar core approaches temperature around log T = 10 K. We observed
a similar trend for electron captures on $^{55}Co$ [26] but the
capture rates of this nucleus were slower than electron capture
rates on $^{56}Ni$. We also noted that capture rates of $^{56}Ni$ is
one order of magnitude faster than that of $^{55}Co$ when the
stellar core shifts from densities ($10^{7}$ to $10^{11}) g/cm^{3}$
at low temperatures (around log T = 7.0).

FFN calculated electron and positron capture rates for nuclei in the
range A = 21-60. The GT contribution to the rate was parameterized
on the basis of the independent particle model and supplemented by a
contribution simulating low-lying transitions. Fig. 4 depicts the
comparison of our electron capture rates with the FFN rates [12] for
densities\textit{ $\rho Y_{e} = 10^{3} g/cm^{3}$} and \textit{ $\rho
Y_{e} = 10^{11} g/cm^{3}$}. At low densities (around \textit{$\rho
$Ye }= $10^{3}g/cm^{3}$) and temperatures (around log T = 9.0), our
electron capture rates for $^{56}Ni$ are in good agreement with FFN
capture rates. As the temperature of the stellar core increases the
FFN gradients becomes steeper. At temperatures log T $>$ 9.5, we
note that the FFN rates are enhanced than our rates. At high
temperatures the probability of occupation of the parent excited
states $({E_{i}})$ increases, FFN did not take into effect the
process of particle emission from excited states (this process is
accounted for in the present pn-QRPA calculations). FFN's parent
excitation energies $({E_{i}})$ are well above the particle decay
channel and partly contribute to the enhancement of their electron
capture rates at higher temperatures.

We also compared our calculation of electron capture rates with
those calculated using large-scale shell model [23]. Fig. 6 in Ref.
[15] compares the two calculations. In order to save space, we
decided not to discuss the comparison in this paper. The
core-collapse simulators should take note of our enhanced electron
capture rates compared to shell model results at
\textit{presupernova temperatures}. (For details we refer to [15].)

One of the channels for the energy release from the star is the
neutrino emission which is mainly from the $e/e^{+}$ capture on
nucleons and $e^{\pm}$ annihilation. Positron capture plays a
crucial role in the dynamics of stellar core. They play an indirect
role in the reduction of degeneracy pressure of the electrons in the
core. Fig. 5 shows our positron capture rates on $^{56}Ni$. We note
that the positron capture rates are very slow as compared to
electron capture on $^{56}Ni$. The positron capture rates enhance as
temperature of the stellar core increases. We also observe that the
positron capture rates are almost the same for the densities in the
range $(10-10^{6})g/cm^{3}$. When the densities increase beyond this
range a decline in the positron capture rate starts. At temperature
log T = 10.5, when the stellar core shifts from density ($10^{7}$ to
$10^{11})g/cm^{3}$, we observe a decline of 3 orders of magnitude in
the positron capture rates.\vskip 0.2in

\begin{center}{\Large\textbf{4. Summary}}\end{center}

We have performed pn-QRPA calculations to determine the presupernova
electron and positron capture rates on $^{56}Ni$ for selected
densities and temperatures from astrophysical point of view.
$^{56}Ni$ is considered to be amongst the most important nuclei for
capturing electrons in the presupernova conditions and core collapse
phase. We have also presented our calculated rates on a finer
temperature-density grid which might prove useful as a test suite
for advanced interpolation routines. Though our centroid is at high
excitation energies in daughter but still our electron capture rates
are enhanced as compared to shell model rates at presupernova
temperatures. Core collapse simulators may find it convenient to
take note of these enhanced capture rates. One of the main reasons
for these enhanced rates is the \textit{microscopic} calculation of
GT strength from the excited states. The pn-QRPA gave us the liberty
of using a large model space of $7\hbar \omega $ and proved to be a
judicious choice for handling excited states in heavy nuclei in the
stellar environment. Table 3 shows our calculations of electron and
positron capture rates on $^{56}Ni$ on a fine grid of
temperature-density scale. (In the online version of this paper
Table 3 is replaced by Table 5 which also contains the
(anti)neutrino energy loss rates.)

Aufderheide et al. [14] reported that the rate of change of
lepton-to-baryon ratio ($\mathop{\Psi }\limits^{.}$) in the stellar
core changes by about 25\% alone due to the electron captures on
$^{56}Ni$. Due to our enhanced electron capture rates in the
presupernova epoch, the core should radiate out more energy by the
process of neutrino emission, keeping the core on a trajectory with
lower temperature and entropy. It is also to be noted that Hix and
colloborators [3] were unable to find an explosion of their
spherically symmetric core collapse simulations. One main reason
pointed out by the authors for this failure was the relatively
suppressed electron capture rates used in their simulations. It
might be interesting to find if our reported rates are in favor of a
(prompt) explosion. \vspace{0.5in}

This work is partially supported by the ICTP (Italy) through the
OEA-project-Prj-16.

\newpage
\begin{figure}[htb]
\includegraphics[width=12cm]{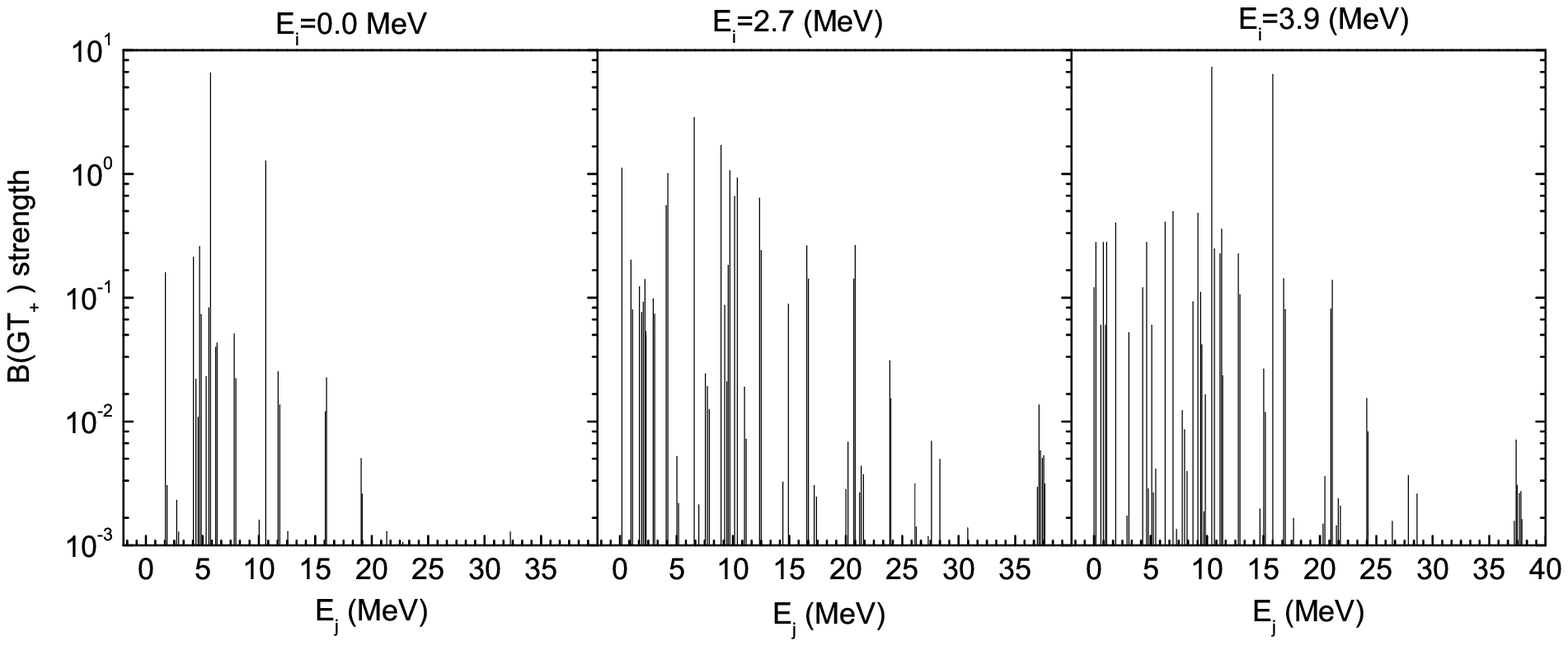}
\caption{Gamow-Teller $(GT_{+})$ strength distributions for
$^{56}Ni$. From left to right, the panels show the $GT_{+}$ strength
for ground, $1^{st}$, and $2^{nd}$ excited states, respectively.
$E_{i}$ $(E_{j})$ represents energy of parent (daughter) states. The
energy scale refers to the excitation energies in the daughter
$^{56}Co$.}
\end{figure}
\begin{figure}[htb]
\includegraphics[width=12cm]{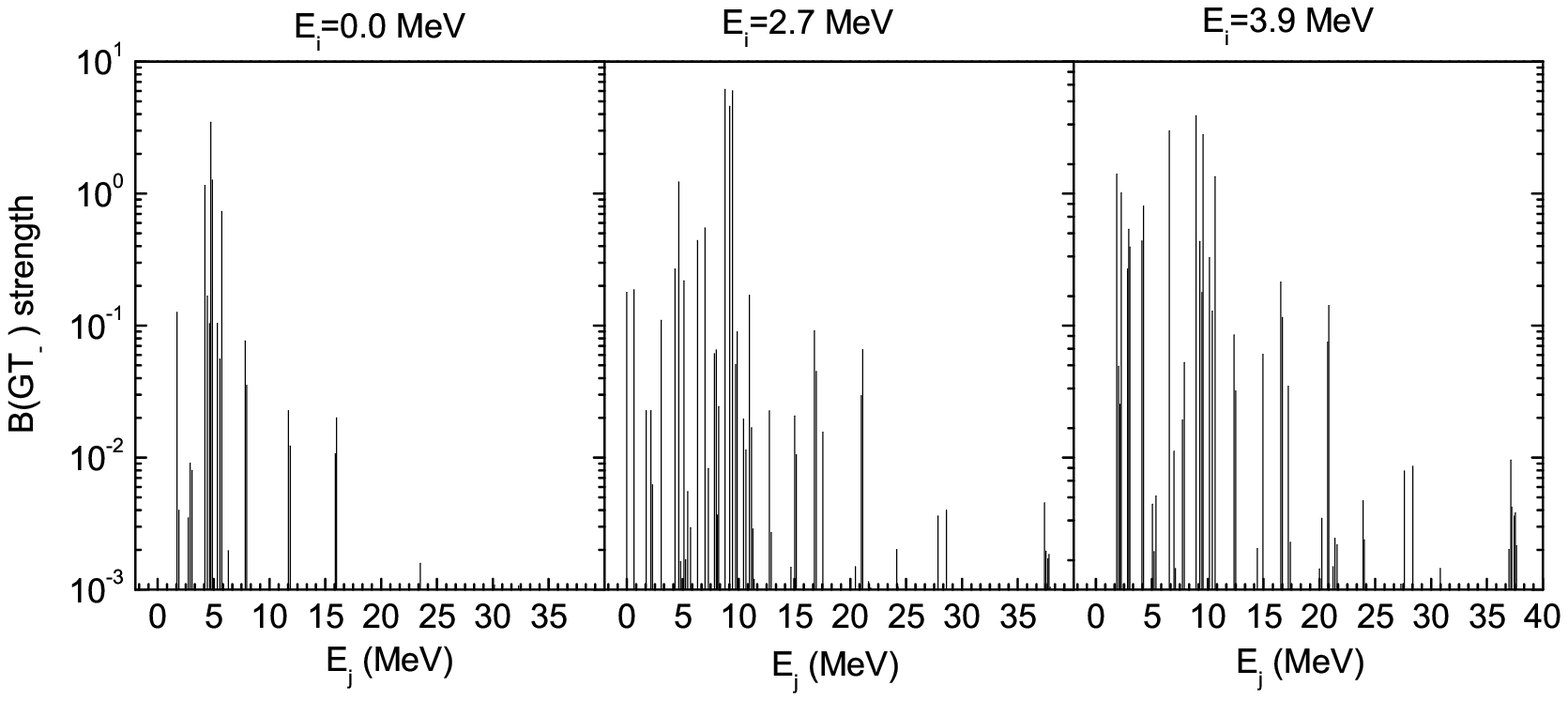}
\caption{Gamow-Teller $(GT_{-})$ strength distributions for
$^{56}Ni$. From left to right, the panels show $GT_{-}$ strength for
ground, $1^{st}$, and $2^{nd}$ excited states, respectively. $E_{i}$
$(E_{j})$ represents energy of parent (daughter) states. The energy
scale refers to excitation energies in the daughter $^{56}Cu$.}
\end{figure}
\begin{figure}[htb]
\includegraphics[width=12cm]{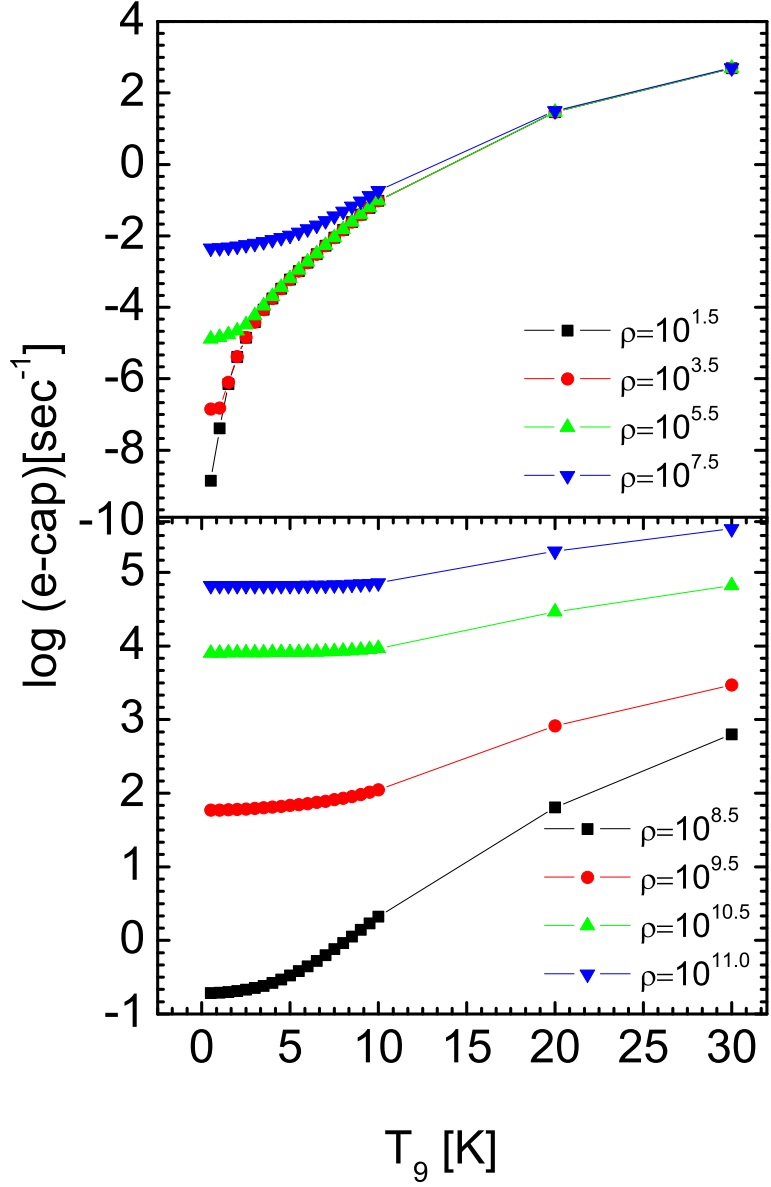}
\caption{Calculated electron captures rates (in logarithmic scale)
on $^{56}Ni$ as function of temperatures for different selected
densities. The densities in the legend are in units of $g/cm^{3}$
whereas $T_{9}$ represents temperature in units of $10^{9}K$.}
\end{figure}
\begin{figure}[htb]
\includegraphics[width=12cm]{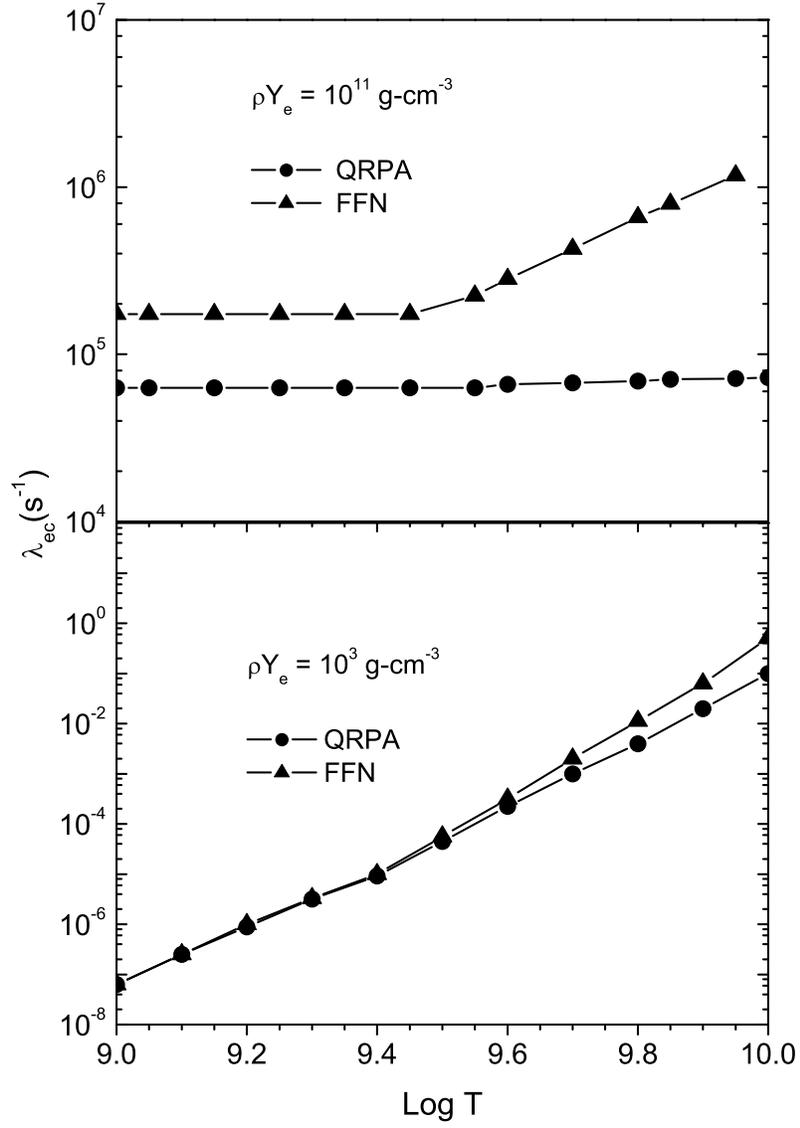}
\caption{Comparison of QRPA electron capture rates with those of
FFN [12] on $^{56}Ni$ as function of temperature. The upper panel is
for density $10^{11}g/cm^{3}$ while the lower panel is for density
$10^{3}g/cm^{3}$.}
\end{figure}
\begin{figure}[htb]
\includegraphics[width=12cm]{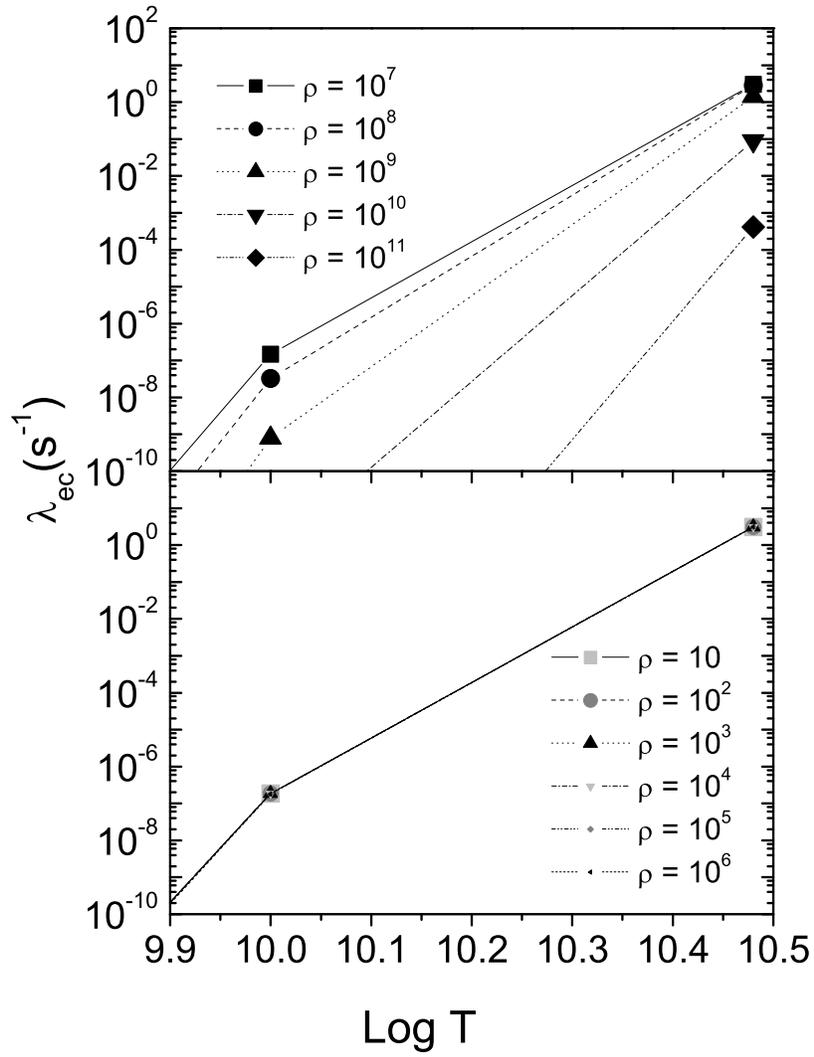}
\caption{Positron captures rates on $^{56}Ni$ as function of
temperatures for different selected densities. The densities in the
legend are in units of $g/cm^{3}$.}
\end{figure}
\clearpage \textbf{Table 1:} Calculated excited states in
parent $^{56}$Ni.\\

\begin{center}
\begin{tabular}{cccccc} \hline\\

 0.00 & 5.23 & 6.01 & 7.29 & 7.76 & 8.80\\
 2.70 & 5.39 & 6.21 & 7.35 & 7.86 & 9.12\\
 3.96 & 5.47 & 6.32 & 7.48 & 8.08 & 9.29\\
 4.97 & 5.68 & 6.44 & 7.53 & 8.31 & 9.71\\
 5.08 & 5.76 & 6.65 & 7.62 & 8.56 & 9.98\\\hline

\end{tabular}
\end{center}
\vspace{1.5in}
\textbf{Table 2a:} Calculated B(GT+) values from ground state in $^{56}$Ni.\\
\begin{center}
\begin{tabular}{cc|cc|cc} \hline\\
Energy(MeV)& B(GT+)& Energy(MeV)& B(GT+)& Energy(MeV)&
B(GT+)\\\hline
1.72 &   1.59E-01 & 4.63 &   1.08E-02 & 6.18 &   3.98E-02\\
1.88 &   3.03E-03 & 4.74 &   2.59E-01 & 6.31 &   4.31E-02\\
2.72 &   2.32E-03 & 4.88 &   7.32E-02 & 7.82 &   5.10E-02\\
2.90 &   1.28E-03 & 5.33 &   2.30E-02 & 7.97 &   2.22E-02\\
4.21 &   2.12E-01 & 5.56 &   8.27E-02 &  10.03&   1.60E-03\\
4.44 &   2.20E-02 & 5.73 &   6.54E+00 &                    \\\hline
\end{tabular}
\end{center}
\vskip 0.2in

\textbf{Table 2b:} Calculated B(GT-) values from ground state in $^{56}$Ni.\\
\begin{center}
\begin{tabular}{cc|cc|cc} \hline\\
Energy(MeV)& B(GT-)& Energy(MeV)& B(GT-)& Energy(MeV)&
B(GT-)\\\hline
1.72 &   1.26E-01  & 4.44 &   1.67E-01 & 5.73 &  7.32E-01\\
1.88 &   3.98E-03  & 4.63 &  1.03E-01  & 6.31 &   1.97E-03\\
2.72 &   3.50E-03  & 4.74 &   3.48E+00 & 7.82 &   7.67E-02\\
2.90 &   9.08E-03  & 4.88 &   1.27E+00 & 7.97 &  3.52E-02 \\
3.08 &   8.00E-03  & 5.33 &  1.04E-01  &                  \\
4.21 &   1.15E+00  & 5.56 &   5.59E-02 &                   \\\hline
\end{tabular}
\end{center}
\clearpage \textbf{Table 3:} Calculated electron and positron
capture rates on $^{56}Ni$ for different selected densities and
temperatures in stellar matter. Aden is log($\rho Y_{e}$) and has
units of $g/cm^{3}$, where $\rho$ is the baryon density and $Y_{e}$
is the ratio of the electron number to the baryon number.
Temperatures ($T_{9}$) are measured in $10^{9}K$ . E-cap and E+cap
are the electron and positron capture rates, respectively. The
calculated electron and positron capture rates are tabulated in
logarithmic (to base 10) scale in units of $sec^{-1}$. In the table,
-100.000 means that the rate is smaller than $10^{-100}$.\\

\begin{center}
\begin{tabular}{cccc|cccc} \hline\\
     Aden & $T_{9}$ & E-cap & E+cap        &     Aden & $T_{9}$ & E-cap & E+cap    \\ \hline
     0.5 &  0.50 &    -9.828 &   -100.000  &     1.5 &  0.50 &     -8.850 &   -100.000  \\
     0.5  & 1.00 &    -7.394 &   -81.298  &      1.5 &  1.00 &    -7.387 &    -81.305  \\
     0.5 &  1.50 &    -6.148  &   -54.201 &      1.5 &  1.50 &    -6.147  &   -54.201  \\
     0.5  & 2.00 &    -5.391 &    -40.556 &      1.5 &  2.00 &   -5.391  &   -40.556  \\
     0.5 &  2.50 &    -4.846 &    -32.317 &      1.5  & 2.50 &    -4.846  &   -32.317  \\
     0.5  & 3.00  &   -4.418 &    -26.790 &      1.5 &  3.00 &     -4.418  &   -26.790  \\
     0.5  & 3.50  &   -4.063 &    -22.817 &      1.5 &  3.50 &   -4.062 &    -22.817 \\
     0.5 &  4.00 &    -3.756 &    -19.818 &      1.5 &  4.00 &    -3.755&   -19.817 \\
     0.5 &  4.50 &     -3.481 &    -17.468 &    1.5 &  4.50 &    -3.481 &    -17.468  \\
     0.5 &  5.00 &     -3.227 &    -15.575 &    1.5 &  5.00 &     -3.226  &   -15.575  \\
     0.5 &  5.50 &    -2.984 &    -14.014  &    1.5 &  5.50 &     -2.983  &   -14.013   \\
     0.5 &  6.00 &    -2.747 &    -12.702 &     1.5 &  6.00 &     -2.746  &   -12.702 \\
     0.5 &  6.50 &    -2.512 &    -11.583 &     1.5 &  6.50 &    -2.512   &  -11.583  \\
     0.5 &  7.00 &   -2.281 &    -10.615   &    1.5 &  7.00 &    -2.280  &   -10.615  \\
     0.5 &  7.50 &    -2.053 &     -9.769  &    1.5 &  7.50 &    -2.053   &   -9.768  \\
     0.5 &  8.00 &     -1.831 &     -9.021 &    1.5 &  8.00 &    -1.831 &     -9.020   \\
     0.5 &  8.50 &    -1.616 &     -8.354  &    1.5 &  8.50 &     -1.616   &   -8.354  \\
     0.5 &  9.00 &    -1.410 &     -7.756  &    1.5 &  9.00 &     -1.409    &  -7.755  \\
     0.5 &  9.50 &    -1.211 &     -7.215  &    1.5 &  9.50 &     -1.211  &    -7.215  \\
     0.5 & 10.00 &     -1.022 &     -6.724  &   1.5 & 10.00 &     -1.021   &   -6.723  \\
     0.5 & 20.00 &      1.464 &     -1.625 &    1.5 & 20.00 &     1.465  &    -1.623   \\
     0.5 & 30.00 &      2.691 &      0.473  &    1.5 & 30.00 &    2.693  &     0.474  \\

     1.0 &  0.50 &    -9.348 &   -100.000  &2.0 &  0.50 &  -8.350 &    -100.000\\
     1.0  & 1.00 &    -7.392 &    -81.300   &2.0 &  1.00 &    -7.370  &   -81.321\\
     1.0 &  1.50 &     -6.148  &   -54.200 &2.0 &  1.50 &    -6.146 &    -54.202  \\
     1.0 &  2.00 &    -5.391  &   -40.556  &2.0  & 2.00 &    -5.390  &   -40.556  \\
     1.0 &  2.50 &    -4.846  &   -32.317 &2.0 &  2.50 &     -4.846  &   -32.317  \\
     1.0 &  3.00 &    -4.418  &   -26.790  &2.0 &  3.00 &     -4.418 &    -26.790  \\
     1.0 &  3.50 &    -4.062  &   -22.817  &2.0 &  3.50 &     -4.062  &   -22.817 \\
     1.0 &  4.00 &    -3.755  &   -19.817  &2.0 &  4.00 &    -3.755 &    -19.817 \\
     1.0 &  4.50 &    -3.481  &   -17.468   &2.0 &  4.50 &    -3.480  &   -17.468 \\
     1.0 &  5.00 &     -3.226  &   -15.575 &2.0 &  5.00 &   -3.226  &   -15.575 \\
     1.0 &  5.50 &    -2.983  &   -14.014  &2.0 &  5.50 &    -2.983 &    -14.013  \\
     1.0 &  6.00 &     -2.746 &    -12.702 &2.0 &  6.00 &     -2.746 &    -12.702  \\
     1.0 &  6.50  &    -2.512  &   -11.583  &2.0 &  6.50 &    -2.512  &   -11.583 \\
     1.0 &  7.00 &    -2.280  &  -10.615   &2.0 &  7.00 &     -2.280  &   -10.615 \\
     1.0 &  7.50 &     -2.053  &    -9.768 &2.0 &  7.50 &    -2.053  &    -9.768\\
     1.0 &  8.00 &     -1.831  &    -9.020 &2.0 &  8.00 &   -1.831 &     -9.020 \\
     1.0 &  8.50 &     -1.616   &   -8.354  &2.0 &  8.50 &     -1.616 &     -8.354 \\
     1.0 &  9.00 &    -1.409   &   -7.756   &2.0 &  9.00 &    -1.409  &    -7.755 \\
     1.0 &  9.50 &    -1.211  &    -7.215   &2.0 &  9.50 &    -1.211  &    -7.215 \\
     1.0 & 10.00 &     -1.021  &    -6.723  &2.0 & 10.00  &   -1.021  &    -6.723 \\
     1.0 & 20.00 &      1.465  &    -1.624  &2.0 & 20.00 &      1.465 &     -1.623 \\
     1.0 & 30.00 &      2.692   &    0.474  &  2.0 & 30.00 &  2.693   &     0.474\\

\end{tabular}
\end{center}
\newpage
\begin{center}
\begin{tabular}{cccc|cccc} \hline\\

     Aden & $T_{9}$ & E-cap  & E+cap         &     Aden & $T_{9}$ & E-cap   & E+cap    \\ \hline
     2.5&   0.50&       -7.850&    -100.000 &  3.5 &  0.50 &     -6.851&   -100.000  \\
     2.5&   1.00&      -7.319 &    -81.373 &   3.5 &  1.00 &    -6.818&    -81.875 \\
     2.5&   1.50 &    -6.143 &    -54.205  &   3.5 &  1.50 &     -6.097&    -54.251  \\
     2.5 &  2.00 &    -5.390 &    -40.557  &   3.5 &  2.00  &   -5.380 &   -40.567 \\
     2.5&   2.50 &    -4.846 &    -32.317  &   3.5 &  2.50 &     -4.842&    -32.321   \\
     2.5 &  3.00 &     -4.418 &    -26.790 &   3.5  & 3.00 &     -4.416&    -26.791  \\
     2.5&   3.50 &    -4.062 &    -22.817  &   3.5 &  3.50  &    -4.061 &   -22.818   \\
     2.5 &  4.00&      -3.755  &   -19.817 &   3.5 &  4.00 &   -3.755 &   -19.818  \\
     2.5 &  4.50&      -3.480 &    -17.468 &   3.5  & 4.50 &     -3.480 &   -17.468  \\
     2.5 &  5.00 &     -3.226 &    -15.575 &   3.5 &  5.00 &   -3.226 &   -15.575  \\
     2.5&   5.50 &     -2.983 &    -14.013 &   3.5&   5.50 &    -2.983 &   -14.014  \\
     2.5&   6.00 &    -2.746 &    -12.702  &   3.5 &  6.00 &   -2.746 &   -12.702  \\
     2.5 &  6.50 &    -2.512  &   -11.583 &    3.5 &  6.50&     -2.512 &   -11.583   \\
     2.5 &  7.00  &   -2.280  &   -10.615  &   3.5&   7.00&      -2.280 &   -10.615 \\
     2.5&   7.50 &    -2.053  &    -9.768  &   3.5 &  7.50&      -2.053 &    -9.768 \\
     2.5 &  8.00 &     -1.831  &    -9.020 &   3.5 &  8.00&      -1.831 &    -9.020  \\
     2.5 &  8.50 &    -1.616  &    -8.354  &   3.5&   8.50 &    -1.616 &    -8.354  \\
     2.5&   9.00 &    -1.409  &    -7.755  &   3.5 &  9.00&    -1.409 &    -7.755  \\
     2.5 &  9.50 &    -1.210  &    -7.215  &   3.5 &  9.50&     -1.210 &    -7.215 \\
     2.5&  10.00 &    -1.021  &    -6.723  &   3.5 & 10.00&      -1.021 &    -6.723  \\
     2.5 & 20.00 &      1.465  &    -1.623 &   3.5 & 20.00 &     1.465 &    -1.623  \\
     2.5 & 30.00 &       2.693  &     0.475&   3.5 & 30.00 &     2.693 &     0.475  \\

     3.0 &  0.50 &    -7.350 &   -100.000 &    4.0&   0.50 &    -6.353 &  -100.000  \\
     3.0 &  1.00 &     -7.165 &    -81.527 &   4.0 &  1.00 &     -6.346 &   -82.350  \\
     3.0 &  1.50 &     -6.132 &    -54.216 &   4.0&   1.50 &    -5.991 &   -54.358  \\
     3.0 &  2.00 &     -5.387 &    -40.559 &    4.0 &  2.00 &     -5.356 &   -40.591  \\
     3.0 &  2.50 &    -4.845  &   -32.318  &   4.0&   2.50 &    -4.834 &   -32.329  \\
     3.0 &  3.00 &     -4.417  &   -26.790  &  4.0 &  3.00 &     -4.412  &  -26.796  \\
     3.0 &  3.50 &     -4.062 &    -22.817 &   4.0 &  3.50 &    -4.059 &   -22.820 \\
     3.0 &  4.00&      -3.755  &   -19.817  &  4.0 &  4.00 &    -3.753  &  -19.819  \\
     3.0 &  4.50 &    -3.480 &    -17.468  &   4.0 &  4.50 &    -3.479 &   -17.469  \\
     3.0&   5.00 & -3.226  &   -15.575      &   4.0 &  5.00 &    -3.225  &  -15.575  \\
     3.0 &  5.50 &     -2.983 &    -14.013  &   4.0&   5.50 &    -2.982 &   -14.014   \\
     3.0 &  6.00 &    -2.746 &    -12.702  &    4.0 &  6.00 &     -2.745 &   -12.702  \\
     3.0 &  6.50 &   -2.512 &    -11.583  &     4.0 &  6.50 &    -2.511  &  -11.583 \\
     3.0 &  7.00 &     -2.280 &    -10.615  &   4.0 &  7.00 &     -2.280  &  -10.615  \\
     3.0 &  7.50 &    -2.053 &    -9.768   &    4.0 &  7.50 &   -2.052  &   -9.768  \\
     3.0 &  8.00 &     -1.831 &    -9.020  &    4.0 &  8.00 &    -1.830  &   -9.020  \\
     3.0 &  8.50 &     -1.616 &    -8.354  &    4.0 &  8.50 &    -1.616 &    -8.354  \\
     3.0&   9.00 &    -1.409&     -7.755  &     4.0 &  9.00 &     -1.409  &   -7.755  \\
     3.0&   9.50 &    -1.210 &    -7.214  &     4.0 &  9.50 &    -1.210  &   -7.215  \\
     3.0&  10.00 &    -1.021 &    -6.723  &      4.0 & 10.00 &     -1.021  &   -6.723  \\
     3.0&  20.00 &     1.465 &    -1.623  &      4.0 & 20.00 &     1.465  &   -1.623  \\
     3.0&  30.00 &     2.693  &    0.475   &     4.0 & 30.00 &     2.693  &    0.475  \\
\end{tabular}
\end{center}
\newpage

\begin{center}
\begin{tabular}{cccc|cccc} \hline\\
Aden & $T_{9}$ & E-cap   & E+cap          &     Aden & $T_{9}$ &
E-cap   & E+cap    \\ \hline
     4.5 &  0.50 &     -5.858 &  -100.000  & 5.5 &  0.50 &     -4.889 &  -100.000 \\
     4.5 &  1.00 &    -5.849 &   -82.855  &  5.5 &  1.00 &     -4.845  &  -83.961 \\
     4.5 &  1.50 &     -5.715 &   -54.636 &   5.5 &  1.50 &     -4.768  &  -55.622  \\
     4.5 &  2.00 &     -5.282 &   -40.666  &  5.5 &  2.00 &     -4.663  &  -41.300  \\
     4.5 &  2.50 &    -4.807 &   -32.357 &    5.5 &  2.50 &     -4.490  &  -32.680  \\
     4.5 &  3.00 &     -4.399 &   -26.808  &  5.5 &  3.00 &     -4.237  &  -26.974  \\
     4.5 &  3.50 &    -4.052 &   -22.827  &   5.5 &  3.50 &    -3.961  &  -22.920  \\
     4.5 &  4.00 &     -3.749 &   -19.823 &   5.5 &  4.00 &     -3.693 &   -19.881  \\
     4.5 &  4.50 &   -3.476 &   -17.472   &   5.5 &  4.50 &     -3.439 &   -17.510  \\
     4.5 &  5.00 &     -3.223 &   -15.577   & 5.5 &  5.00 &     -3.197 &   -15.604   \\
     4.5 &  5.50 &     -2.981 &   -14.015  &  5.5 &  5.50 &     -2.962  &  -14.035   \\
     4.5 &  6.00 &     -2.744 &   -12.703  &  5.5 &  6.00 &   -2.730  &  -12.718   \\
     4.5 &  6.50 &    -2.511 &  -11.584   &   5.5 &  6.50 &    -2.500  &  -11.595 \\
     4.5 &  7.00 &     -2.279 &   -10.615  &  5.5 &  7.00 &    -2.271  &  -10.624   \\
     4.5 &  7.50 &     -2.052 &    -9.769  &  5.5 &  7.50 &     -2.045  &   -9.776  \\
     4.5 &  8.00&      -1.830 &    -9.021  &  5.5 &  8.00 &    -1.824  &   -9.026  \\
     4.5 &  8.50&      -1.615 &    -8.354  &  5.5 &  8.50 &    -1.611  &   -8.359  \\
     4.5 &  9.00 &    -1.408 &    -7.756  &   5.5 &  9.00 &    -1.405  &   -7.760  \\
     4.5 &  9.50 &     -1.210 &    -7.215  &  5.5 &  9.50 &    -1.207  &   -7.218  \\
     4.5&  10.00 &    -1.020 &    -6.723  &   5.5 & 10.00 &    -1.018  &   -6.726  \\
     4.5 & 20.00 &      1.465  &   -1.623  &  5.5 & 20.00 &     1.466  &   -1.624  \\
     4.5 & 30.00&      2.693  &    0.475  &   5.5 & 30.00 &     2.693  &    0.475  \\

     5.0 &  0.50 &     -5.372 &  -100.000 &   6.0 &  0.50 &    -4.377 &  -100.000 \\
     5.0 &  1.00 &      -5.348 &   -83.381 &  6.0 &  1.00 &    -4.321 &   -84.685 \\
     5.0 &  1.50 &    -5.269  &  -55.092  &   6.0 &  1.50 &   -4.246  &  -56.229 \\
     5.0 &  2.00 &    -5.071 &   -40.879  &   6.0 &  2.00 &    -4.158  &  -41.847  \\
     5.0 &  2.50 &    -4.723 &   -32.442  &    6.0 &  2.50 &    -4.058 &   -33.133  \\
     5.0 &  3.00 &     -4.359 &   -26.849  &   6.0 &  3.00 &    -3.930 &   -27.292  \\
     5.0 &  3.50 &    -4.030 &   -22.849   &   6.0 &  3.50 &    -3.762 &   -23.125 \\
     5.0 &  4.00 &     -3.735  &  -19.837   &  6.0 &  4.00 &    -3.564 &   -20.014  \\
     5.0 &  4.50 &     -3.467 &   -17.481   &   6.0 &  4.50 &    -3.352 &   -17.599  \\
     5.0 &  5.00 &    -3.217 &   -15.584   &   6.0 &  5.00 &   -3.136 &   -15.667   \\
     5.0 &  5.50 &     -2.977 &   -14.020   &  6.0 &  5.50 &    -2.918 &   -14.080  \\
     5.0 &  6.00 &    -2.741 &   -12.707   &   6.0 &  6.00 &    -2.697 &   -12.752 \\
     5.0 &  6.50 &    -2.508 &   -11.586   &   6.0 &  6.50 &    -2.474  &  -11.621  \\
     5.0 &  7.00 &    -2.277 &   -10.618   &   6.0 &  7.00 &    -2.250  &  -10.645  \\
     5.0 &  7.50 &     -2.050 &    -9.770   &  6.0 &  7.50 &    -2.029  &   -9.792  \\
     5.0 &  8.00 &     -1.829 &    -9.022  &  6.0 &  8.00 &   -1.811  &   -9.040 \\
     5.0 &  8.50 &    -1.614 &    -8.355   &  6.0 &  8.50 &    -1.600  &   -8.370 \\
     5.0 &  9.00 &    -1.407 &    -7.757   &  6.0 &  9.00 &    -1.395  &   -7.769  \\
     5.0 &  9.50 &    -1.209 &    -7.216    & 6.0 &  9.50 &    -1.199  &   -7.226  \\
     5.0 & 10.00 &   -1.020 &    -6.724    &  6.0 & 10.00 &    -1.011  &   -6.733   \\
     5.0 & 20.00 &    1.465  &   -1.623   &   6.0 & 20.00 &     1.467  &   -1.624  \\
     5.0 & 30.00 &      2.693  &    0.475   & 6.0 & 30.00 &     2.693  &    0.474  \\
\end{tabular}
\end{center}
\newpage

\begin{center}
\begin{tabular}{cccc|cccc} \hline\\

Aden & $T_{9}$ & E-cap   & E+cap          &     Aden & $T_{9}$ &
E-cap  & E+cap    \\ \hline
     6.5 &  0.50 &    -3.785 &  -100.000  &  7.5 &  0.50 &     -2.345 &  -100.000 \\
     6.5 &  1.00 &     -3.741 &   -85.731  & 7.5 &  1.00 &     -2.332 &  -89.832  \\
     6.5 &  1.50 &   -3.678 &   -57.012  &  7.5 &  1.50 &     -2.310 &   -59.827 \\
     6.5 &  2.00 &     -3.605 &   -42.513  & 7.5 &  2.00 &     -2.281 &  -44.709  \\
     6.5 &  2.50 &    -3.526 &   -33.729  & 7.5 &  2.50 &     -2.245 &   -35.571 \\
     6.5 &  3.00 &   -3.442 &   -27.818  &  7.5 &  3.00 &     -2.205 &   -29.432 \\
     6.5 &  3.50 &    -3.349 &   -23.561  & 7.5 &  3.50 &    -2.159 &   -25.011  \\
     6.5 &  4.00 &     -3.240 &   -20.351  & 7.5 &  4.00 &    -2.108 &   -21.667  \\
     6.5 &  4.50 &     -3.109 &   -17.851   & 7.5 &  4.50 &    -2.049 &   -19.042  \\
     6.5 &  5.00 &    -2.955 &   -15.854   &  7.5 &  5.00 &     -1.980 &   -16.921   \\
     6.5 &  5.50 &    -2.781 &   -14.220  &  7.5 &  5.50 &    -1.899  &  -15.169  \\
     6.5 &  6.00 &    -2.593 &   -12.859   & 7.5 &  6.00 &     -1.804 &   -13.695 \\
     6.5 &  6.50 &   -2.393 &   -11.704   &  7.5 &  6.50 &    -1.696 &   -12.436 \\
     6.5 &  7.00 &    -2.186 &   -10.711   & 7.5 &  7.00 &     -1.576 &   -11.348 \\
     6.5 &  7.50 &    -1.977 &    -9.845   &  7.5 &  7.50 &     -1.446 &   -10.397  \\
     6.5 &  8.00 &  -1.769 &    -9.083   &  7.5 &  8.00 &    -1.309 &    -9.560  \\
     6.5 &  8.50 &    -1.565 &    -8.406   & 7.5 &  8.50 &     -1.167 &    -8.818  \\
     6.5 &  9.00 &    -1.366 &    -7.799    & 7.5 &  9.00 &     -1.022 &    -8.155 \\
     6.5 &  9.50 &    -1.175  &   -7.251   & 7.5 &  9.50 &     -0.876 &    -7.560 \\
     6.5 & 10.00 &    -0.990  &   -6.754  & 7.5 & 10.00 &     -0.731 &    -7.022 \\
     6.5 & 20.00  &   1.469  &   -1.627   & 7.5 & 20.00 &      1.502 &    -1.661 \\
     6.5 & 30.00&     2.694  &    0.474    & 7.5 & 30.00 &      2.704 &     0.464 \\

     7.0 &  0.50 &    -3.101&   -100.000  & 8.0&   0.50 &    -1.550 &  -100.000  \\
     7.0 &  1.00 &     -3.074&    -87.347  & 8.0 &  1.00 &     -1.543 &   -93.578  \\
     7.0 &  1.50 &     -3.034 &   -58.140  & 8.0 &  1.50 &     -1.532 &   -62.344  \\
     7.0 &  2.00 &     -2.983 &   -43.412  & 8.0 &  2.00 &     -1.516 &   -46.619  \\
     7.0 &  2.50 &    -2.925 &   -34.499  & 8.0 &  2.50 &     -1.496 &   -37.121  \\
     7.0 &  3.00 &    -2.863 &   -28.506   & 8.0 &  3.00 &     -1.471 &  -30.746   \\
     7.0 &  3.50 &     -2.797 &   -24.185   & 8.0 &  3.50 &    -1.440 &   -26.160  \\
     7.0 &  4.00 &     -2.726  &  -20.914  &  8.0 &  4.00 &   -1.402 &   -22.694  \\
     7.0 &  4.50 &     -2.646  &  -18.347  & 8.0 &  4.50  &   -1.355 &   -19.977   \\
     7.0 &  5.00 &     -2.552  &  -16.278  & 8.0 &  5.00 &    -1.298 &   -17.784  \\
     7.0 &  5.50 &     -2.442  &  -14.575  & 8.0 &  5.50 &     -1.231 &   -15.973  \\
     7.0 &  6.00 &     -2.311  &  -13.151  & 8.0 &  6.00 &   -1.152 &  -14.450   \\
     7.0 &  6.50 &     -2.162  &  -11.943   & 8.0 &  6.50 &     -1.063 &   -13.148  \\
     7.0 &  7.00 &     -1.997  &  -10.906  & 8.0 &  7.00 &    -0.966 &   -12.020  \\
     7.0 &  7.50 &     -1.821  &  -10.005  & 8.0 &  7.50 &    -0.861 &  -11.032  \\
     7.0 &  8.00 &     -1.640  &   -9.215   & 8.0 &  8.00 &     -0.752 &   -10.159  \\
     7.0 &  8.50 &     -1.457  &   -8.516   &  8.0 &  8.50 &    -0.641  &   -9.380  \\
     7.0 &  9.00 &     -1.276  &   -7.892   &  8.0 &  9.00 &    -0.528  &   -8.681  \\
     7.0 &  9.50 &    -1.098  &   -7.330   &   8.0 &  9.50 &   -0.415 &    -8.049  \\
     7.0 & 10.00 &    -0.925   &  -6.822   &   8.0 & 10.00&     -0.302  &   -7.475  \\
     7.0 & 20.00 &     1.477  &   -1.635   &   8.0 & 20.00 &     1.580  &   -1.742  \\
     7.0 & 30.00 &     2.696  &    0.471   &   8.0 & 30.00 &     2.727  &    0.440 \\
\end{tabular}
\end{center}
\newpage

\begin{center}
\begin{tabular}{cccc|cccc} \hline\\

     Aden & $T_{9}$ & E-cap   & E+cap          &     Aden & $T_{9}$ & E-cap   & E+cap   \\ \hline
     8.5 &  0.50 &    -0.714 &  -100.000  & 9.5 &  0.50  &      1.770 &  -100.000 \\
     8.5 &  1.00 &    -0.709 &   -99.150  & 9.5 &  1.00  &     1.772 &  -100.000   \\
     8.5 & 1.50  &  -0.701  &  -66.073    & 9.5 &  1.50  &     1.776 &   -79.658 \\
     8.5 &  2.00  &  -0.689  & -49.429    & 9.5 &  2.00  &    1.780  &  -59.635  \\
     8.5 &  2.50 &    -0.671  &  -39.384  & 9.5 &  2.50  &     1.786  &  -47.565   \\
     8.5 &  3.00 &   -0.648  &  -32.647   & 9.5 &  3.00  &     1.793  & -39.482    \\
     8.5 &  3.50  &   -0.617 &   -27.805  & 9.5 &  3.50  &     1.802 &   -33.681   \\
     8.5 &  4.00 &  -0.579  &  -24.149    & 9.5 &  4.00 &       1.811 &   -29.308   \\
     8.5 &  4.50 &    -0.532 &   -21.285  & 9.5 & 4.50  &    1.822  &  -25.889   \\
     8.5 &  5.00 &    -0.478 &   -18.977  & 9.5 &  5.00 &      1.833 &   -23.138   \\
     8.5 &  5.50 &   -0.417 &   -17.073   & 9.5 &  5.50 &   1.846  &  -20.874    \\
     8.5 &  6.00 &    -0.350 &   -15.473  & 9.5 &  6.00 &     1.860 &   -18.975   \\
     8.5 &  6.50 &     -0.278 &   -14.107 & 9.5 &  6.50 &     1.875 &   -17.357    \\
     8.5 &  7.00 &    -0.200 &   -12.925  & 9.5 &  7.00 &     1.892 &   -15.961  \\
     8.5 &  7.50 &    -0.119 &   -11.891  & 9.5 &  7.50 &    1.911 &  -14.742    \\
     8.5 &  8.00 &    -0.034 &  -10.977  & 9.5 &  8.00 &    1.932  &  -13.668    \\
     8.5 &  8.50 &    0.054  & -10.162    & 9.5 &  8.50 &    1.956 &   -12.712    \\
     8.5 &  9.00 &     0.142 &    -9.430  & 9.5 &  9.00 &     1.983  &  -11.856     \\
     8.5 &  9.50 &     0.232 &    -8.768  & 9.5 &  9.50 &    2.013  & -11.083      \\
     8.5 & 10.00 &     0.322 &   -8.166   & 9.5 & 10.00 &    2.046  &  -10.383     \\
     8.5 & 20.00 &   1.806  &   -1.977    & 9.5 & 20.00 &     2.913 &    -3.212     \\
     8.5&  30.00  &    2.800 &     0.364  & 9.5&  30.00 &    3.470 &    -0.341     \\

     9.0 &  0.50  &     0.385 &  -100.000   &10.0 &  0.50 &     2.905 &  -100.000 \\
     9.0 &  1.00  &     0.393 &  -100.000   &10.0 &  1.00 &      2.906  & -100.000 \\
     9.0 &  1.50  &     0.407 &   -71.571   &10.0 &  1.50 &     2.907 &   -91.541 \\
     9.0 &  2.00  &     0.425 &   -53.563    &10.0 &  2.00 &     2.908  &  -68.552 \\
     9.0 &  2.50  &   0.447  & -42.701     &10.0 &  2.50 &     2.910  &  -54.703 \\
     9.0 &  3.00  &    0.474  &  -35.421     &10.0 &  3.00 &   2.912 &   -45.435  \\
     9.0  & 3.50  &     0.504 &   -30.193  &10.0 &  3.50 &   2.915  &  -38.788   \\
     9.0 &  4.00  &    0.537  &  -26.249   &10.0 &  4.00 &     2.918 &   -33.782  \\
     9.0 &  4.50  &    0.572  &  -23.162   &10.0 &  4.50 &    2.921 &   -29.870  \\
     9.0 &  5.00  &    0.610  &  -20.677   & 10.0  & 5.00 &     2.925 &   -26.726 \\
     9.0 &  5.50  &    0.649 &   -18.629   &10.0 &  5.50&      2.929 &   -24.141  \\
     9.0 &  6.00  &    0.690 &   -16.910    & 10.0  & 6.00&     2.934  &  -21.975  \\
     9.0  & 6.50  &     0.734 &   -15.444   &10.0 &  6.50&      2.940  &  -20.131  \\
     9.0 &  7.00  &    0.779 &   -14.177   & 10.0 & 7.00 &    2.946   & -18.542   \\
     9.0 &  7.50  &   0.827  &  -13.070   &10.0 &  7.50 &     2.954 &   -17.156   \\
     9.0 &  8.00  &    0.878  &  -12.093   &10.0 &  8.00 &    2.963 &   -15.936  \\
     9.0 &  8.50  &    0.932  &  -11.222   &10.0 &  8.50 &   2.974 &   -14.852  \\
     9.0 &  9.00  &    0.989  &  -10.442   &10.0 &  9.00 &     2.988 &   -13.882  \\
     9.0&   9.50 &  1.048   &  -9.736     &10.0 &  9.50 &    3.003 &   -13.008   \\
     9.0 & 10.00  &    1.110  &   -9.096   &10.0 & 10.00 &    3.021 &   -12.216  \\
     9.0 & 20.00 &    2.274  &   -2.477  &10.0&  20.00 &    3.657 &    -4.204  \\
     9.0 & 30.00  &    3.014    &  0.144   &10.0 & 30.00 &     4.099  &   -1.064  \\
\end{tabular}
\end{center}
\newpage

\begin{center}
\begin{tabular}{cccc|cccc} \hline\\

    Aden & $T_{9}$ & E-cap   & E+cap          &     Aden & $T_{9}$ & E-cap   & E+cap    \\ \hline

    10.5 &  0.50 &    3.905 &  -100.000  &11.0 &  0.50&     4.821 &  -100.000 \\
    10.5 &  1.00 &    3.905 &  -100.000  & 11.0 &  1.00 &   4.821 &  -100.000 \\
    10.5 &  1.50 &   3.906 &  -100.000 &11.0 &  1.50 &    4.821 &  -100.000 \\
    10.5 &  2.00 &   3.906 &   -81.642  &11.0 &  2.00 &   4.821 &  -100.000 \\
    10.5 &  2.50 &    3.907 &   -65.179  &11.0 &  2.50 &    4.821  &  -80.554 \\
    10.5 &  3.00 &   3.908 &   -54.168  &11.0 &  3.00 &    4.822  &  -66.983 \\
    10.5 &  3.50 &    3.909 &   -46.277  & 11.0 &  3.50 &    4.822 & -57.263   \\
    10.5 &  4.00 &   3.910 &   -40.338  &11.0 &  4.00&     4.822 & -49.953 \\
    10.5 &  4.50 &3.911   & -35.701    &11.0 &  4.50&     4.823 & -44.251 \\
    10.5 &  5.00 &   3.912  &  -31.977  & 11.0 & 5.00&    4.823  &  -39.674  \\
    10.5 &  5.50 &   3.914  &  -28.918  & 11.0 &  5.50 &   4.824 &  -35.917 \\
    10.5 &  6.00 &   3.916  &  -26.357  &11.0 &  6.00 &   4.825 &  -32.775\\
    10.5 &  6.50 &  3.918   & -24.180  &11.0 &  6.50 &  4.826 &   -30.107\\
    10.5 &  7.00 &   3.921  &  -22.305  &11.0 &  7.00 & 4.828  &   -27.811  \\
    10.5 &  7.50 &   3.925  &  -20.672  &11.0 &  7.50 &  4.830 &   -25.813 \\
    10.5 &  8.00 &   3.930  &  -19.235   & 11.0 &  8.00 &  4.833 &   -24.057 \\
    10.5 &  8.50 &   3.936  &  -17.960   &11.0 & 8.50  & 4.838  &   -22.501  \\
    10.5 &  9.00 &  3.944  &  -16.821  &11.0 &  9.00 & 4.843  &   -21.112  \\
    10.5&   9.50 &   3.954  &  -15.796  & 11.0 &  9.50 &  4.851 &   -19.863 \\
    10.5&  10.00 &   3.965  &  -14.868  &11.0 & 10.00 &  4.860 &   -18.734\\
    10.5&  20.00 &   4.465  &   -5.582  & 11.0 & 20.00 &  5.291 &    -7.552 \\
    10.5 & 30.00  &   4.821  &   -2.039  &11.0 & 30.00  & 5.592 &    -3.392 \\ \hline
\end{tabular}
\end{center}

\end{document}